\documentclass[twocolumn,amsmath,amssymb]{jpsj3}
\usepackage{amsmath,amsfonts,amsthm,amssymb,slashbox} 
\usepackage{times}
\usepackage{color}
\usepackage{graphicx}
\usepackage{dcolumn}
\usepackage{bm}
\usepackage{ulem}
\definecolor{green}{rgb}{0,0.5,0.1}
\definecolor{blue}{rgb}{0,0,0.8}

\newcommand{\eqsa}[1]{\begin{eqnarray} #1 \end{eqnarray}}
\newcommand{\eqwd}[1]{
\onecolumn 
\twocolumn} 

\definecolor{green}{rgb}{0,0.5,0.1}
\definecolor{blue}{rgb}{0,0,0.8}
\begin{document}

\title{Electron Correlation Effects on Topological Phases}

\author{Masatoshi Imada, Youhei Yamaji and Moyuru Kurita
}
\inst{Department of Applied Physics, University of Tokyo, Hongo, Bunkyo-ku, Tokyo 113-8656 
}
\abst{Topological insulators are found in materials that have elements with strong spin orbit interaction. However, electron Coulomb repulsion also potentially generates the topological insulators as well as Chern insulators by the mechanism of spontaneous symmetry breaking, which is called topological Mott insulators. The quantum criticality of the transition to the topological Mott insulators from zero-gap semiconductors follows unconventional universality distinct from the Landau-Ginzburg-Wilson scenario. On the pyrochlore lattice, the interplay of the electron correlation and the spin orbit interaction provides us in a rich phase diagram not only with simple topological insulators but also with Weyl semimetal and topologically distinct antiferromagnetic phases. Magnetic domain wall of the all-in-all-out type antiferromagnetic order offers a promising candidate of magnetically controlled  transport, because, even when the Weyl points disappears, the domain wall maintains robust gapless excitations with Fermi surfaces around it embedded in the bulk insulator and bears uniform magnetization simultaneously.  The ingap state is protected by a mechanism similar to the solitons in polyacetylene. Puzzling experimental results of pyrochlore iridates are favorably compared with the prediction of the domain wall theory.         
}

\kword{topological insulator, strongly correlated systems, spin-orbit interaction, pyrochlore iridates}

\maketitle

\section{Introduction}

Existence of topological insulators were first theoretically predicted by Kane and Mele~\cite{KaneMele} for systems with strong spin-orbit interaction and experimentally proven to exist in HgTe/(Hg,Cd)Te quantum wells based on the theoretical prediction.\cite{Konig} It has a bulk charge excitation gap similarly to the band insulator, while in contrast to the ordinary band insulators, they accompany protected metallic surfaces described by Dirac-type Fermions.
They have attracted much interest firstly because they are new states of matter and secondly because they are potential candidates of future spintronics by utilizing the spin-dependent and protected electronic transport on the surface.   

\begin{figure}[htb]
	\centering
	\includegraphics[width=8cm]{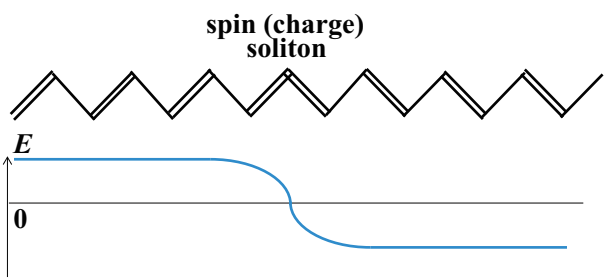}
\caption{(Color online) Schematics of polyacetylene soliton with inversion of the gap $E$}
	\label{polyacetylene}
\end{figure}

The protection of the gapless metallic surface is explained by the band inversion of the gap state between two topologically distinct phases. One illuminating way of intuitively understanding the emergence of the gapless state is found in the example of solitons in  polyacetylene.\cite{SuSchriefferHeeger} The two topologically distinct domains in the opposite side of a charge/spin soliton illustrated in Fig.~\ref{polyacetylene} can be viewed as having the $\pi$-phase shifted gaps. To connect these gaps with the opposite sign continuously within the real gap through the soliton, one needs to go through a zero-gap state, which generates ingap states with a gapless excitation.  

Emergence of topological states in two-dimensional systems were found in the quantum Hall state under the strong magnetic field with the time-reversal symmetry breaking. Even without external magnetic fields, if microscopically alternating fluxes penetrating the lattice exist, Haldane~\cite{Haldane} showed the existence of a quantum Hall state on the honeycomb lattice, where the gap opens in bulk and the gapless and protected charge current is induced on the edge as we discuss below. This quantum Hall state is called Chern insulator and are discussed in a different context of the flux state as a theoretical model of the cuprate high-$T_c$ superconductors.\cite{Chakravarty,Varma} The topological insulator is viewed as a phase similar to the quantum Hall state but by replacing the charge loop current with the spin loop current with preserved time reversal symmetry. 

The topological insulators so far identified experimentally always contain elements with strong spin-orbit interaction such as Hg and Au, where the electron correlation is believed to be relatively weak.
In addition, it was shown that the metallic surface is topologically protected against the electron correlation effect at least when the Coulomb interaction can be regarded weak and the electron localization is suppressed by the absence of the backscattering.\cite{backscattering}

Nevertheless, a quite different possibility of realizing the topological insulators induced by the pure electron correlation effect without the spin orbit interaction was theoretically proposed by the mechanism of a spontaneous symmetry breaking caused by the mean-field decoupling of the intersite Coulomb interaction. This mechanism is based on the fact that the direct intersite Coulomb interaction of the form 
$V_{ij}n_in_j$ can be decoupled to the product of two terms, each of which is proportional to the form of the spin-orbit-interaction in the Fock decoupling procedure as we discuss later. Here, $n_i$ is the electron density at the $i$-th site and the  amplitude of intersite Coulomb interaction between the Wannier orbitals on the $i$-th and $j$-th sites is $V_{ij}$.
It was first examined by the Hartree-Fock approximation for the honeycomb lattice with next-neighbor Coulomb interaction stronger than the nearest-neighbor and onsite interactions.\cite{Raghu}
Similar proposals have also been made for the kagom\'{e},\cite{cit:PRB82_075125} diamond~\cite{cit:PRB79_245331} and pyrochlore lattices.\cite{cit:PRL103_206805,cit:PRB82_085111}
The possibility of a topological insulator was proposed for the three-dimensional (3D) pyrochlore lattice for more realistic values of interactions, where a nearest-neighbor interaction substantially weaker than the onsite repulsion stabilizes the topological insulator  in the moderate correlation regime against the competition with various antiferromagnetic and charge ordered states.\cite{Kurita10}  

When the spin orbit interaction and electron correlation are both substantial, in several cases, they are competing each other.  One typical competition is expected in the competition of the topological insulator and magnetic order.  Since the magnetic order such as the antiferromagnetic order breaks the time reversal symmetry, it destroys the topological insulator. On the other hand, when the gap of the topological insulator is large, it cannot be destroyed by a weak electron correlation. Therefore, a phase transition is expected from the topological phase to the magnetically ordered phase.  Even when the magnetic order is stabilized, it was recently proposed that the topological effect remains, where Weyl Fermions emerges in the bulk with arc-like Fermi surface on the surface.\cite{Wan11} On the other hand, it was shown that the charge or (or charge density wave) can coexist with the topological insulator.\cite{Kurita10}

\begin{figure}
	\centering
	\includegraphics{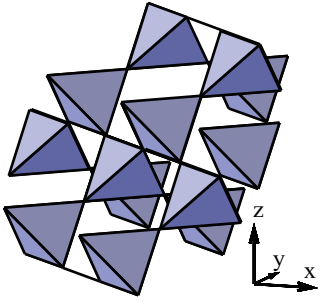}
	\caption{(Color online) Pyrochlore lattice  }
	\label{pyrochlore_lattice}
\end{figure}

In this review, we focus on the role of electron correlation effects in physics of topological phases, particularly by taking an example of electrons on the pyrochlore lattice, whose lattice structure is the stacking of the elementary tetrahedron as shown in Fig. \ref{pyrochlore_lattice}.

\section{Topological Mott insulator} 

\begin{figure}
	\centering
	\includegraphics[width=2.5cm]{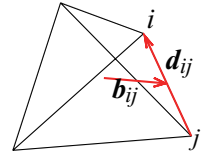}
\caption{(Color online) Unit cell with graphical definitions of $\bm{b}_{ij}$ and $\bm{d}_{ij}$.}
	\label{pyrochlore_unitcell}
\end{figure}

We start from a single-band Hubbard-type model Hamiltonian on the pyrochlore lattice defined by
\begin{equation}
H = H_{0} + H_{\rm SO} + U\sum_{i}n_{i\uparrow}n_{i\downarrow} + V\sum_{\langle ij \rangle }n_{i}n_{j},
\label{Hamiltonian}
\end{equation}
\begin{equation}
	H_0 = -\sum_{\langle ij \rangle \sigma}t_{i,j}(c^{\dagger}_{i\sigma}c_{j\sigma}+h.c.), \label{tight-binding}
\end{equation}
and 
\begin{equation}
	H_{\mathrm{SO}} = \sqrt{2}\lambda\sum_{\langle ij \rangle \alpha\beta}\left(i c_{i\alpha}^{\dagger}\frac{\bm{b}_{ij}\times\bm{d}_{ij}}{|\bm{b}_{ij}\times\bm{d}_{ij}|}\cdot\bm{\sigma}_{\alpha\beta}c_{j\beta} + h.c. \right), 
	\label{spin-orbit}
\end{equation}
where $t_{ij}$, $U$, $V$ and $\lambda$ are the electronic transfer for the bond $(i,j)$, onsite Coulomb repulsion, the nearest neighbor Coulomb interaction and the spin orbit coupling strength, respectively. Here, $n_{i\sigma} = c_{i\sigma}^{\dagger}c_{i\sigma}$, $n_{i} = n_{i\uparrow}+n_{i\downarrow}$ and $c_{i\sigma}(c_{i\sigma}^{\dagger})$ is the electron annihilation (creation) operators on the $i$-th site with spin $\sigma$ and $\langle ij \rangle$ is the summation over the nearest neighbor pairs. The vector $\bm{b}_{ij}$ bridges the center of a tetrahedron to the midpoint of the bond $\langle ij \rangle$ constituting the tetrahedron edge and $\bm{d}_{ij}$ is the vector connecting the site $j$ to $i$ as illustrated in Fig. \ref{pyrochlore_unitcell}.
Since each bond participates in forming only a single tetrahedron, $\bm{b}_{ij}$ is uniquely determined when a bond $\langle ij \rangle$ is specified.

Since the unit cell contains 4 sites, The tight binding Hamiltonian $H_0$ generates four bands among which two upper bands are degenerate and completely flat (dispersionless). One of the other two bands touches the flat bands at the $\Gamma$ point and disperses quadratically from there.  This quadratic touching band is completely filled at half filling and the flat bands are empty, so that the system is a zero-gap semiconductor and behaves as a semimetal.

In real materials with pyrochlore structure, they have orbital degrees of freedom near the Fermi level such as $t_{2g}$ manifold of the 5$d$ elements in Ir pyrochlore oxides, $R_2$Ir$_2$O$_7$, where $R=$Pr, Nd, Sm, Eu, Gd, .... In these cases, 6-fold degenerate $t_{2g}$ manifold (including the spin degrees of freedom) is located near the Fermi level, separated from the $e_g$ manifold by the crystal field splitting. In the presence of the spin-orbit interaction, these 6-fold degenerate orbitals are split into the lower 4-fold orbitals with the total angular momentum $J=3/2$ and the upper two orbitals with $J=1/2$, where $J=1/2$ band becomes half filling for $R_2$Ir$_2$O$_7$.  Then the single band description by eq.(\ref{Hamiltonian}) is justified when the $J=1/2$ band is isolated near the Fermi level. 

\begin{figure}
\centering
\includegraphics[width=5cm]{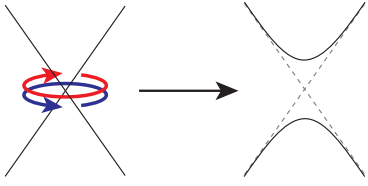}
\caption{
(Color online) Illustration for degeneracy lifting  of clockwise and anticlockwise motion of electrons at a band-crossing point, which generates the gap by the spin orbit interaction.
The loop current in this momentum space is equivalent to that in the real space shown in Fig.~\ref{spin_loop_current}}. 
\label{clockwise-anticlockwise}
\end{figure}

\begin{figure}[htb]
\begin{center}
	\includegraphics[width=2.5cm]{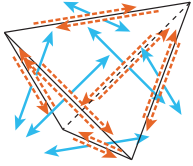}
\end{center}
	\caption{(Color online) Illustration of the spin loop current: Dashed (red) arrows along bonds show the current directions of the spins in the direction (solid (blue) arrows) attached to the current arrows. Along each bond, the opposite spins flow in the opposite directions that makes spin current, while the spin direction of the flow is always perpendicular to the bond. }
	\label{spin_loop_current}
\end{figure}

When the spin-orbit interaction $\lambda$ is positive, a gap immediately opens at the quadratic touching point that leads to an insulator. The topological index indicates that this insulator is a strong topological insulator.  A gap opens at the band crossing point induced by the spin-orbit interaction, that lifts the degeneracy of the clockwise and anticlockwise spin current state existing when $\lambda=0$ as in Fig.\ref{clockwise-anticlockwise}. This lift means that a microscopic spin current flows in the elementary tetrahedron as in Fig.\ref{spin_loop_current}. 
The spin loop current is defined by
\begin{eqnarray}
	\zeta_s &=& \frac{i}{2} \sum_{\alpha,\beta} \langle c^{\dagger}_{j\beta}c_{i\alpha} \rangle \frac{\bm{b}_{ij}\times \bm{d}_{ij}}{|\bm{b}_{ij}\times \bm{d}_{ij}|}\cdot \bm{\sigma}_{\beta\alpha},
	\label{zeta_s}
\end{eqnarray}
which creates time-reversal-symmetric topological insulators, if $\zeta_s$ does not depend on the bond $(i,j)$.

However, the uniform spin loop current cancels because the spin current along each bond has different spin direction which becomes zero when summed up in the tetrahedron. Therefore, the spin current in the bulk cannot be detected, while if there exists a surface, this cancellation becomes incomplete and the net spin current flows with gapless excitations. This surface spin current is topologically protected.  Note that the quantum Hall (Chern) insulator is understood from the microscopic charge loop current instead of the spin loop current in the case of the topological insulator. 
The charge-orbital current is defined by
\begin{eqnarray}
	\zeta_c &=& \frac{i}{2} \sum_{\sigma}
	\langle c^{\dagger}_{j\sigma}{c}_{i\sigma} \rangle
	\left[ \frac{\bm{b}_{ij}\times \bm{d}_{ij}}{|\bm{b}_{ij}\times \bm{d}_{ij}|} \right]_{z}.
	\label{zeta_c}
\end{eqnarray}
The Chern insulator can be realized by applying either uniform external magnetic field or the alternating flux penetrating the two-dimensional plane.\cite{Haldane} 

When $\lambda$ is negative, which is the realistic case of $R
_2$Ir$_2$O$_7$, the lift of degeneracy occurs between two flat bands, while the quadratic band touching at the  $\Gamma$ point between one of the originally flat band and the band with the originally quadratic dispersion remains at the Fermi level. Therefore, the semimetallic bulk phase continues for negative $\lambda$.
The point $\lambda=0$ is a quantum critical point, where a semimetal undergoes a metal-insulator transition to a topological insulator when we cross from the $\lambda <0$ region to the $\lambda>0$ region.

When the Coulomb interactions $U$ and $V$ are switched on, an emergent feature appears.\cite{Kurita10} This can be first viewed by the mean-field approximation.
To construct the mean-field picture, we decouple the on-site interaction according to
\begin{eqnarray}
	n_{i\uparrow}n_{i\downarrow} &\approx& n_{i\uparrow}\langle n_{i\downarrow}\rangle + \langle n_{i\uparrow}\rangle n_{i\downarrow} - \langle n_{i\uparrow}\rangle \langle n_{i\downarrow}\rangle \notag \\
	&-& c_{i\uparrow}^{\dagger} c_{i\downarrow}\langle c_{i\downarrow}^{\dagger} c_{i\uparrow}\rangle - \langle c_{i\uparrow}^{\dagger} c_{i\downarrow}\rangle c_{i\downarrow}^{\dagger} c_{i\uparrow} \\
	&+& \langle c_{i\uparrow}^{\dagger} c_{i\downarrow}\rangle \langle c_{i\downarrow}^{\dagger} c_{i\uparrow} \rangle , \label{U_Fock}
\end{eqnarray}
and the nearest-neighbor interaction as
\begin{eqnarray}
	n_{i}n_{j}&\approx&n_{i}\langle n_{j}\rangle + \langle n_{i}\rangle n_{j} - \langle n_{i}\rangle \langle n_{j}\rangle \notag \\ 
	&-& \sum_{\alpha\beta} (c_{i\alpha}^{\dagger}c_{j\beta}\langle c_{j\beta}^{\dagger}c_{i\alpha}\rangle  
 + \langle c_{i\alpha}^{\dagger}c_{j\beta}\rangle c_{j\beta}^{\dagger}c_{i\alpha} 
\notag \\
    &-&\langle c_{i\alpha}^{\dagger}c_{j\beta}\rangle \langle 
c_{j\beta}^{\dagger}c_{i\alpha}\rangle ). \label{V_Fock}
\end{eqnarray}
We note that when the system is a semimetal, where no symmetry is broken, $\langle n_{i\uparrow}\rangle = \langle n_{i\downarrow}\rangle , \langle c_{i\uparrow}^{\dagger}c_{j\uparrow}\rangle = \langle c_{i\downarrow}^{\dagger}c_{j\downarrow}\rangle \neq 0$, and $\langle c_{i\uparrow}^{\dagger}c_{j\downarrow}\rangle = \langle c_{i\downarrow}^{\dagger}c_{j\uparrow}\rangle = 0$ are satisfied.
The emergent spin loop current (eq.(\ref{zeta_s})) is generated from the Fock decoupling in eq. (\ref{V_Fock}) as 
\begin{eqnarray}
	\langle c^{\dagger}_{j\beta}c_{i\alpha} \rangle &=& g\sigma_{0\alpha\beta} -i\zeta_s \frac{\bm{b}_{ij}\times \bm{d}_{ij}}{|\bm{b}_{ij}\times \bm{d}_{ij}|}\cdot \bm{{\sigma}}_{\alpha\beta}.
	\label{V_Fock_zeta}
\end{eqnarray}
Charge loop current eq.(\ref{zeta_c}) can also be generated similarly.
When the spontaneous symmetry breaking with the nonzero order parameter $\zeta_s$ or $\zeta_c$ emerging at $\lambda=0$ is called a topological Mott insulator (TMI). 
 
\begin{figure}
	\centering
	\includegraphics[width=6.5cm]{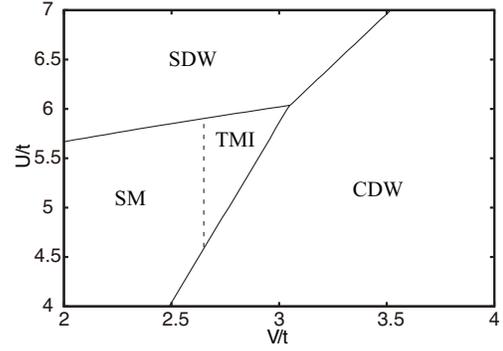}
	\caption{Phase diagram of extended Hubbard model on pyrochlore lattice. The solid (dashed) lines indicate first-order (continuous) transitions. SM, TMI, CDW, and SDW denote semimetal, topological Mott insulator, charge density wave and spin density wave, respectively.}
	\label{phase_diagram}
\end{figure}

\begin{figure}[htb]
\begin{center}
	\includegraphics[width=0.25\textwidth]{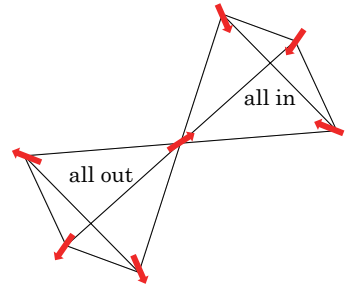}
\end{center}
	\caption{(Color online) All-in all-out type antiferromagnetically ordered state illustrated for the unit cell of two neighboring tetrahedrons.}
	\label{all_in_all_out}
\end{figure}

The phase diagram obtained from the Hartree Fock approximation with the above Fock decoupling is shown in Fig.~\ref{phase_diagram} for $\lambda=0$.  The strong topological insulating phase becomes the ground state sandwiched by the all-in-all-out type antiferromagnetic ordered (SDW)  phase illustrated in Fig.~\ref{all_in_all_out} and the charge ordered (CDW) phase representing the unit tetrahedron with two charge rich and two charge poor sites regularly stacking. 
Therefore, the spontaneous symmetry breaking to the spin loop current phase illustrated in Fig.~\ref{spin_loop_current} is certainly an emergent and realistic possibility in the  intermediate correlation regime in the absence of the spin-orbit interaction at least in the Hartree-Fock level.  

\begin{figure}[htb]
\centering
\includegraphics[width=8.0cm]{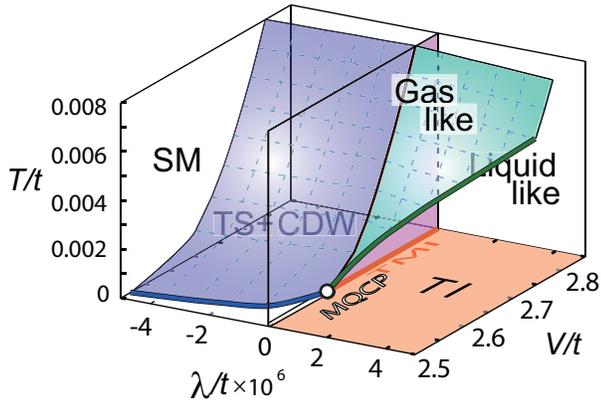}
\caption{
(Color online)
Phase diagram of pyrochlore lattice in the plane of $V$, $\lambda$ and temperature $T$ at $U/t=5.5$.\cite{KuritaPRB}
In the region $\lambda>0$ at temperature $T=0$, the strong topological insulator is stabilized for all $V\ge 0$.
At $T>0$, a light green surface separates a gas-like and liquid-like topological semiconductors (TS). This surface of the first-order transition terminates on the Ising critical (bold green) line, which further terminates at the marginal quantum critical point (MQCP)(white circle).
In the absence of the spin-orbit interaction, $\lambda=0$, a quantum phase transition from semimetal (SM) for small $V$ to topological Mott insulator (TMI) (dark orange line) occurs at the MQCP, $V=V_{c}\simeq 2.62$t at $T=0$ (white circle).
The MQCP shows a new universality beyond the Landau-Ginzburg-Wilson scheme. 
When $\lambda < 0$, SM at small $V$ undergoes a first-order transition (the light blue surface) to a topological semiconductor coexisting with charge order (TS+CDW). When $U$ becomes large TS+CDW is replaced with the all-in all-out order (as one can speculate from Fig.~\ref{phase_diagram}, where the Weyl semimetal also emerges as we discuss later.
The TMI coexiting with charge order (TMI+CDW) at $\lambda\le 0, T=0$ crossovers to TS+CDW at nonzero temperatures.
It terminates at a quantum critical line (dark blue). 
The MQCP also terminates two quantum critical lines at $T=0$, one along $\lambda=0, V<V_{c}$ (bold white line) and the other (the dark blue line) representing that between TS+CDW and SM. 
}
\label{Phase_Diagram_V_lambda}
\end{figure}
The interplay of the electron correlation and the spin orbit interaction is seen in the phase diagram for $U/t=5.5$ in the plane of $V$ and $\lambda$ in Fig.\ref{Phase_Diagram_V_lambda}.
This phase diagram reveals various interesting emergent phenomena arising from the topological phases and the electron correlation effect.  The phase diagram is basically categorized into three parts: One is the strong topological insulator phase at $\lambda>0$. The second is the line along $\lambda=0$. The third is the region $\lambda<0$, where several interesting topologically nontrivial phases emerge combined either with the all-in-all-out type magnetic phase or with charge ordered phase when the interaction gets large. 

The strong topological insulator at $\lambda>0$ and $T=0$ continues to be adiabatically connected to the liquid-like topological semiconductor at nonzero temperature while it undergoes a transition to a gas-like topological semicondustor phase if $V/t$ gets large. This first-order gas-liquid type transition terminates at the critical line depicted as the green bold curve. 
When $\lambda$ changes from positive to negative for weak electron correlation, a continuous quantum phase transition from the topological insulator to a semimetal takes place at $T=0$, at a quantum critical line as illustrated as the white bold line along $\lambda=0$ and $T=0$. This quantum critical line terminates at the marginal quantum critical point (MQCP) at $V_c$(white circle), beyond which the topological Mott insulator (TMI) (bold orange line) is stabilized even without the spin orbit interaction ($\lambda=0$) for appropriate choices of $U$ and $V$ as one sees in Fig.~\ref{phase_diagram}. The topological insulator phase emerges because of the spontaneous symmetry breaking of the SU(2) symmetry breaking of the direction of the spin for the spin loop current in Fig.~\ref{spin_loop_current}. When one crosses from positive $\lambda$ to negative $\lambda$ across the topological Mott insulator, it undergoes a first-order transition into the topologically nontrivial phase either with the charge order or all-in-all-out magnetic order. This first-order transition continues to nonzero temperature and terminates at the critical line (black bold curve). Along this critical line, the critical temperature $T_c$ is lowered with decreasing $V$ and eventually $T_c$ becomes zero at MQCP.  

In the $\lambda<0$ region, the TMI phase switches to topological charge ordered phase or the all-in-all-out order phase depending on $U$ and $V$.  This $T=0$ phase again undergoes a first-order transition into semimetallic phase through the shaded (blue) sheet. This sheet of the first-order transition terminates at the quantum critical line (bold (blue) curve) at $T=0$. When the electron correlation is weak, the semimetal is the stable phase.     
This rich phase diagram reveals a typical interplay of electron correlation and the topological phases.  
 
\begin{figure}[htb]
\centering
\includegraphics[width=6.0cm]{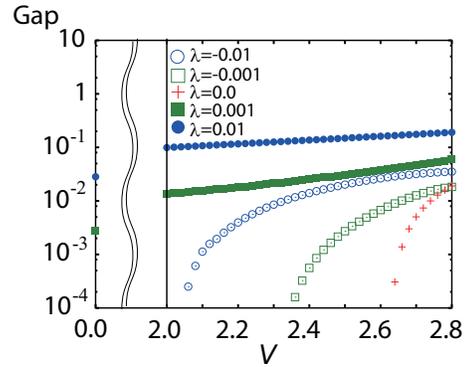}
\caption{(Color online)
Enhancement of bulk charge gap by synergy of electron correlation and spin orbit interaction $\lambda$. In comparison to the gap in the absence of the interaction, the nearest neighbor interaction $V$ largely enhances the bulk gap for the same value of $\lambda$.   
}
\label{Enhance}
\end{figure}

If the spin-orbit interaction is combined with the electron correlation, the gap is synergetically enhanced as we see in Fig.~\ref{Enhance}. It opens a possibility of enhancing the charge gap, which helps in extracting the contribution of the surface/edge transport.

\section{Unconventional quantum criticality}
Let us forcus on the quantum criticality around MQCP. The critical exponents of MQCP was examined on the Hartree Fock level.  It was clarified that the order parameter $\zeta$ grows as $\zeta \propto (V-V_c)^{\beta}$ as $\beta=2$.  At $V=V_c$ with increasing $\lambda$, it grows as $\zeta\propto \lambda^{1/\delta}$ with $\delta=3/2 $. At $\lambda=0$ and $V=V_c$, the order parameter susceptibility $\chi$ diverges as $\chi\propto T^{-\gamma}$ with $\gamma=1$.
This is accounted for by the free energy in the expansion
\begin{eqnarray}
	f[\zeta]-f[0] = -\lambda \zeta + a\zeta^2 +b_{\pm}f_{\rm s}[\zeta] +({\rm higher\ order}),
	\label{Eq.5}
\end{eqnarray}
for small $\zeta$ close to MQCP.
Here $a$ and $b_{\pm}$ are constants.
We add a spin-orbit coupling or magnetic flux $\lambda$ conjugate to $\zeta$, as a straightforward analogue of magnetic fields in magnetic phase transitions.
In addition to the regular term proportional to $a$ that vanishes when MQCP is approached, a singular part $f_{\rm s}[\zeta] \propto |\zeta|^{5/2}$ emerges at $T=0$. 
The coefficient $b_{+}$ for $\zeta > 0$ is not necessarily equal to $b_-$ for $\zeta < 0$. 
The expansion eq.(\ref{Eq.5}) is only piecewise analytic separately in $\zeta>0$ and $\zeta<0$ with a nonanalyticity at $\zeta=0$, originating from topological nature of this transition, in contrast to analytic expansions assumed in the framework of Landau-Ginzburg and Wilson.
Although it satisfies the conventional scaling law $\beta (\delta-1)= \gamma$, it does not follow the requirement $\delta\ge 3$ and $\beta \le 1/2$ satisfied in the usual Landau-Ginzburg type scheme, because the expansion contains nonanalytic term $\propto \zeta^{2.5}$, which originates from the topological transition of the Fermi surface from a semimetal to an insulator, where the density of states has a singularity. Therefore MQCP represents an unconventional quantum critical point, which is indeed the starting point of the quantum critical line, where the topological transition of the Fermi surface continues at $T=0$. 

The critical exponents may be subject to quantitatively change when we consider fluctuations beyond the mean-field level.  However, the topological character of the transition combined with the spontaneous symmetry breaking is retained and enforces the unconventional character of the quantum criticality to be robust.  A related case of an antiferromagnetic transition at a finite $U$ for the simple Hubbard model on the honeycomb lattice was suggested to have only a minor modification from the mean-field value $\beta=1$ to $\beta=0.86$ in the estimate by the quantum Monte Carlo studies.\cite{Assaad}  This indeed violates the conventional Landau-Ginzburg value $\beta <1/2$.

\section{Weyl semimetal and magnetic domain wall}
The region $\lambda<0$ is important, because experimentally, many of the pyrochlore compounds with strong spin-orbit interactions such as the pyrochlore iridates $A_2$Ir$_2$O$_7$ are expected to belong to the region of negative $\lambda$. In the region $\lambda<0$, the topological Mott insulating phase at $\lambda=0$ tends to be
replaced either with the magnetic order or charge order, when electron correlation is large as we see in Fig.~\ref{Phase_Diagram_V_lambda}.  In fact, for relatively large $V/U$, the topological Mott insulator stabilized at $\lambda=0$  is replaced with the charge order coexisting with the topological phase as we see in Fig.~\ref{Phase_Diagram_V_lambda}.
This coexistence is allowed because the charge order does not break the time reversal symmetry and is not severely destructive to the topological phase.

\begin{figure}[htb]
\begin{center}
	\includegraphics[width=0.28\textwidth]{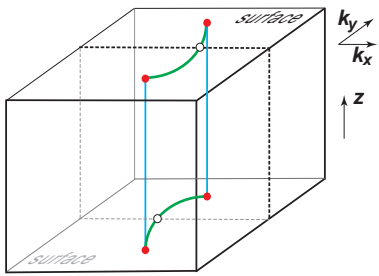}
\end{center}
	\caption{(Color online) Schematics of Weyl points located on $(k_x,k_y)$ momentum plane along the vertical bold (light blue) lines ending on the surface at filled (red) circles. (Note that the $z$ axis is the real space coordinate.) A cross section (sheet surrounded by dashed lines) between one side of the Weyl point is a Chern insulator and it has a gapless point (open circles) on the top and bottom surfaces. Tracing the open circle by moving the cross section, Fermi arcs (bold green curves) appear on the top and bottom surfaces.\cite{Wan11}  }
	\label{WeylArc}
\end{figure} 
On the other hand, the all-in all-out type antiferromagnetic order illustrated in Fig.~\ref{all_in_all_out} is stabilized when $V/U$ is relatively small.  Since it breaks the time reversal symmety, the topological phase could immediately be destroyed and could be replaced with the topologically trivial all-in-all-out phase.  However, Wan {\it et al.}
\cite{Wan11} have clarified that the zero-gap semiconductors with the quadratic touching of the dispersion at the $\Gamma$ point is split into a pair of Weyl points moved from the $\Gamma$ point, where instead of the four component degeneracy including spin at $\lambda=0$, a two-fold degenerate band crossing with gapless dispersions emerges at each Weyl point.

At the Weyl point, the gapless dispersion exists both in the bulk and the surface. 
It was shown that a two-dimensional cross section that makes both of the split two Weyl points into one side of the coross section is a two-dimensional Chern insulator so that the edge of this cross section has gapless excitation as shown in Fig.~\ref{WeylArc}.\cite{Wan11}  This gapless point at the edge forms a line as the trace when one moves the cross section between the two Weyl points, that generates a gapless Fermi surface (actually Fermi line). This trace bridges the two Weyl points. Note that the cross section in the other side of the Weyl points is not topological and the gapless excitation is absent. Therefore, this gapless structure at the surface is called arc, where the Fermi line emerges only as a truncated one which terminates at the Weyl points. 

When the magnetic ordered moment grows, the split Weyl points 
makes a pair annihilation so that the Fermi arc shrinks and disappears as we explain later. After the pair annihilation, no gapless excitations look remaining and the system appears to turn into a trivial insulator both in the bulk and surface. The existence of the Weyl points is limited to very narrow region very close to the semimetal-insulator transition.

\begin{figure}[htb]
\begin{center}
	\includegraphics[width=0.28\textwidth]{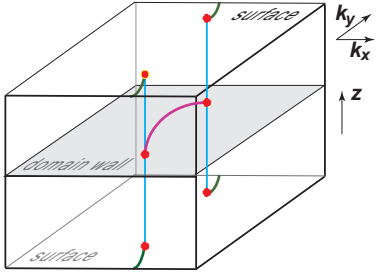}
\end{center}
	\caption{(Color online) Schematics of Weyl points located along the vertical bold (light blue) lines, which form Fermi arc (bold (purple) curve) on the domain wall (shaded middle sheet) in a path  complementary to the arcs on the surfaces (bold (green) curves on the top and bottom surfaces). (Note that we have shifted the Brillouin zone from Fig.~\ref{WeylArc}, so that in this illustration the Fermi arcs on the surface cross the Brillouin zone boundary.) With the growth of the all-in-all-out magnetic moment, the Weyl point easily disappears in pair at the Brilloiun zone boundary here, which results in the disappearance of the arcs on the surfaces as well.  However, the arc on the domain wall extends and makes a closed large Fermi line. \cite{Yamaji13} }
	\label{WeylArcDomainwall}
\end{figure}

However, it was revealed that this wide all-in-all-out phase with the absence of the Weyl points is not a simple insulator.\cite{Yamaji13} This is because instead of the surface, on the magnetic domain wall that separates the two degenerate phases of the all-in-all out and say, all-out-all-in order induces emergent gapless (metallic) two-dimensional sheet, as illustrated in Fig.~\ref{WeylArcDomainwall}.   Here, we review recent studies~\cite{Yamaji13} on  physics of the domain wall and its interesting connection to the emergent transport properties by studying low-energy effective models for the domain wall.

\subsection{ Low-energy effective model for domain wall physics}
\label{Low-energy model of domain wall}
The pyrochlore lattice contains 4 sites in a unit cell and eq.~(\ref{Hamiltonian}) involves 8 component Fermions including spins. Out of 8-component Fermions, the quadratic touching dispersions at the Fermi level contain 4 components, which constitute a 4-component
effective
hamiltonian with the cubic symmetry. This is a variation of the Luttinger hamiltonian\cite{Luttinger56,Murakami04,Moon12} derived from the $\vec{k}\cdot\vec{p}$ perturbation theory for semiconductors if $U=V=0$. 

The low-energy bands of this 4-component Hamiltonian are quadruply degenerate 
at the crystallographic $\Gamma$-point 
$(0,0,0)$,
and form a quadratic band crossing of the bands.

Low-energy physics of the hamiltonian (\ref{Hamiltonian}) for $\lambda < 0$ involves excitations 
around the Weyl point when all-in/all-out orders are formed but weak.
Physics of Weyl electrons can be
captured by mean-field decouplings of the short-ranged Coulomb repulsion $U$ into the mean field for the all-in-all-out type magnetic order, $m$.

\begin{figure}[htb]
\begin{center}
	\includegraphics[width=0.2\textwidth]{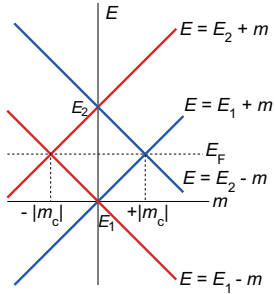}
\end{center}
	\caption{(Color online) Schematics of level splitting of the low-energy states
near the Fermi level. Each of two doubly degenerate states ($E_1$ and $E_2$) 
at $m=0$ splits at nonzero $m$. The blue (dark) lines around the Fermi level $E_F$ is the levels of a domain-wall state confined in the side of the
positive magnetization ($m>0$), while the red (gray) lines around the Fermi level are confined in the opposite $m < 0$ side of the domain wall. The doubly degenerate low-energy excitations in the $m>0$ side (blue (dark) lines) and those for the $m<0$ side (red (gray) lines) are interchanged each other.}
	\label{Gapcross}
\end{figure}

By a small but finite mean field $m$, the 4-fold degeneracy at the $\Gamma$-point
is lifted, and 8 doubly-degenerate Weyl points appear instead at 
$\vec{k}=\vec{k}_{\rm Weyl}\simeq \sqrt{|m|/2t}(\pm 1, \pm 1, \pm 1)$. 
When the 4-component $\vec{k}\cdot\vec{p}$ hamiltonian
generates a doubly-degenerate Weyl point
out of  the 4 components,
the other
two components are gapped and does not enter the low-energy physics as we see in Fig.~\ref{Gapcross}. Then, around the Weyl points, the low-energy effective hamiltonian is 
indeed reduced to a Weyl hamiltonian describing 3D massless Fermions.

Eight  Weyl points become  far apart with the growing $m$. Further increase in $m$ to the order of $t$ results in the 
reassembly of the eight Weyl points into other part of the Brillouin zone and eventually 
are annihilated in pair actually at the four crystallographic L-points,
$\vec{k}=\vec{k}_{\rm L}=(\pm \pi/4a, \pm \pi/4a, \pm\pi/4a)$,
at the boundary of the Brillouin zone.
These appearance (pair creation) and disappearance (pair annihiration) of the Weyl points can then be studied by the  $\vec{k}\cdot\vec{p}$-perturbation theory of the Luttinger Hamiltonian around the $\Gamma$-point and the L-points.

Let us focus on the two cases, the first case where the Weyl points are close to the $\Gamma$ point and the other, close to the L points. 
For $m>0$, we write the Weyl hamiltonian near the $\Gamma$ (L) point as $\hat{h}_{\Gamma\vec{k}_{\rm Weyl}}^{(+)}$ ($\hat{h}_{{\rm L}\vec{k}}^{(+)}$).
Surprisingly, when we change the sign of $m$ with keeping its amplitude,
the gapped two-component hamiltonian for $m>0$, $\hat{h}_{\Gamma\vec{k}_{\rm Weyl}}^{(-)}$ ($\hat{h}_{{\rm L}\vec{k}}^{(-)}$) describes low-energy
Weyl electrons while $\hat{h}_{\Gamma\vec{k}_{\rm Weyl}}^{(+)}$ ($\hat{h}_{{\rm L}\vec{k}}^{(+)}$) becomes gapped
components. So the role is interchanged depending on whether $m>0$ or $m<0$.


For instance, let us consider  the case of  the Weyl points 
$\vec{k}_{\rm Weyl}=\pm \kappa_W (1,1,1)$ with $\kappa_W=\sqrt{|m|/2t}$
for $|m|/t\ll 1$,
and a surface or domain wall perpendicular to $(0,+1,-1)$
(written as $(01\overline{1})$ surface or domain wall).
We assume a periodic boundary condition in the plane direction parallel to the domain wall, which makes the treatment simple and the dependence on the plane direction is characterized by the wavenumber $k_{\parallel}$. The spatial dependence perpendicular to the domain wall is described by taking a real-space new coordinate $X$ in the direction $(0,+1,-1)$ and accordingly, an oblique coordinate $Y$ and $Z$ within the plane parallel to the domain wall may be introduced.  
The corresponding reciprocal momentum coordinate is taken as $(\kappa_X,\kappa_Y,\kappa_Z)$~\cite{Yamaji13}

Then the resultant Dirac hamiltonian around the  $\Gamma$ point is obtained as
$\hat{h}_{\Gamma \vec{k}_{\rm Weyl}}^{(+)}$ and $\hat{h}_{\Gamma \vec{k}_{\rm Weyl}}^{(-)}$ 
for $m > 0$ and $m < 0$, respectively, as
\begin{eqnarray}
\hat{h}_{\Gamma\vec{k}_{\rm Weyl}}^{(\pm)}(-i\partial_X, \delta\kappa_Y, \kappa_Z ; X)
=&&
 h_0 (\delta\kappa_Y, \kappa_Z)\hat{\sigma}_0
 \nonumber
 \\
 +
 h_x (\kappa_Z)\hat{\sigma}_x
 +
 h_y (-i\partial_X) \hat{\sigma}_y
&+&h_z^{(\pm)} (X) \hat{\sigma}_z,
\label{DiracGamma}
\end{eqnarray}
where we
used the new
momentum frame $(\kappa_X, \kappa_Y=\kappa_W+\delta\kappa_Y,\kappa_Z)$
and replaced $\kappa_X$ with $-i\partial_X$.
The Weyl points are projected to $(\delta\kappa_Y , \kappa_Z)=(0,0)$.
Here coefficients in the Dirac hamiltonian (\ref{DiracGamma})
are derived from the original hamiltonian (1), by using the low-energy Luttinger hamiltonian, as
$h_0 = -4t\kappa_W (3\delta\kappa_Y+\kappa_Z)$,
$h_x = 4t\kappa_W \kappa_Z$, $h_y=4t\sqrt{3}\kappa_W i \partial_X$,
and $h_z^{(\pm)} = m(X)\mp |m|$.

This two-component 1D
Dirac equation,
$\hat{h}^{(+)}_{\Gamma\vec{k}_{\rm Weyl}}
\vec{\psi}(X)
 =E
\vec{\psi}(X)
$
describes bound states on the surface or domain walls
by introducing suitable $X$-dependent ``mass" terms $m(X)$.\cite{JackiwRebbi76,FanZhang}
Here, 
the all-out (all-in) domain is described by $m(X)=+|m|$ ($m(X)=-|m|$).
If $|m|$ is large enough,
the Weyl points are annihilated in pair and the bulk system becomes a trivial
magnetic insulator.
Therefore, 
the mass term, $m(X)=
  |m| \theta ( - X )
  -|m| \theta ( X) $ describes magnetic domain wall at $X=0$, 
while a surface between a ``vacuum" ($X<0$)
and the bulk ($X>0$)
can be described by
$
  m(X)=
  |M| \theta ( -X)
  +|m| \theta ( X)$ with $|M|\gg |m|$.

As the order parameter $m$ develops, the two Weyl points
starting as $\vec{k}=\pm|\kappa_W|(1,1,1)$ come closer,
and, finally, are annihilated in pair at the L-point $\vec{k}_{\rm L}=(\pi/4a,\pi/4a,\pi/4a)$.
Around the L-point, the pair of the two-component Dirac hamiltonian
is given as
\begin{equation}
  \hat{h}^{(\pm)}_{{\rm L}\vec{k}_{\rm L}}=
   h_x (\kappa_Z) \hat{\sigma}_x
   +
   h_y (-i\partial_X, \kappa_Z) \hat{\sigma}_y
   +
   h_z^{(\pm)} (X) \hat{\sigma}_z,
   \label{ChiralDiracL}
\end{equation}
where the coefficients of the Pauli matrices $h_x$, $h_y$ are linear functions
of their arguments, and $h_z^{(\pm)} = -m(X)\pm |m|$.
The above Dirac hamiltonians (\ref{ChiralDiracL})
do not depend on $\delta\kappa_Y$, where
$(\kappa_Y,\kappa_Z)=(\pi+\delta\kappa_Y,\kappa_Z)$
, and
the pair-annihilation point is given by $(\delta\kappa_Y,\kappa_Z)=(0,0)$.
When $\kappa_Z=0$,
eq. (\ref{ChiralDiracL}) possesses chiral symmetry
with a chiral operator $\hat{\sigma}_x$



\subsection{Effective one-dimensional Dirac equation}
\label{Section4-2}
If the periodicity along the domain walls is preserved,
the existence of the
ingap states is protected
by the chiral symmetry~\cite{RyuSchnyder,Teo} of the Dirac hamiltonian
at a pair-annihilation point $(\pi, 0)$

The essential physics of these ingap states
is understood from a simplified model 
describing 1D weak Chern insulators,\cite{RyuSchnyder,Teo} where the 
1D Dirac equation  is given as
\eqsa{
\left\{
\left[ 
\pm \alpha (1-\cos \kappa)
-m(X)\right]\hat{\sigma}_z
 +vi\hat{\sigma}_y \partial_X
\right\}
\vec{\psi}(\vec{X})
 =
 E
\vec{\psi}(\vec{X}).
\label{ChD2}
}
Here,
$\kappa$ represents the degrees of freedom of $\kappa_Y$ and $\kappa_Z$.
In eq.(\ref{ChD2}), the emergence of two ``Weyl" points is correctly reproduced as the solution of  $\alpha (1-\cos \kappa)-m(X)=0$. 

\begin{figure}[htb]
\centering
\includegraphics[width=9.0cm]{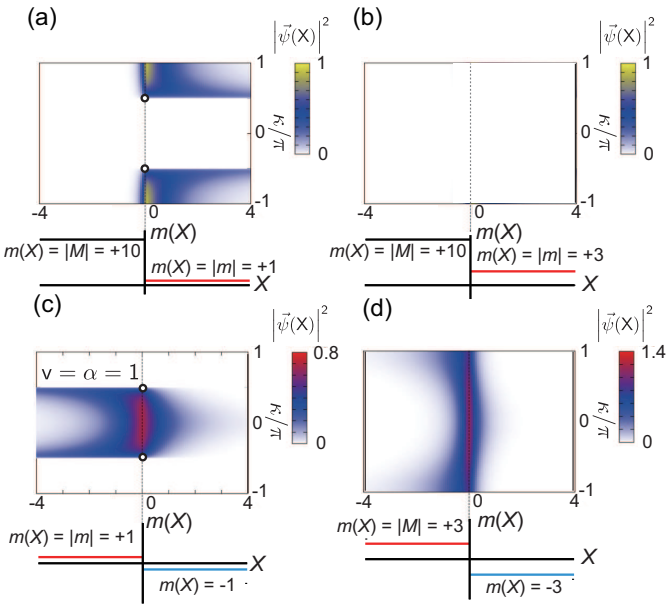}
\caption{
(Color online) Squared wavefunction amplitudes for solutions of effective 1D chiral Dirac equations (\ref{ChD2}).
Surface sandwiched by vacuum (left) and bulk (right), where small $m(=1)$ retains the two Weyl points and arc in (a) and they disappear for large $m(=3)$ in (b). The vacuum is mimicked by large $m(=10)$.  On the other hand, the arc running at the magnetic domain wall represented by $m(x)=m_0(\theta (+X)-\theta (-X))$ for small $m_0(=1)$ in (c) is extended to form full Fermi line for a large $m_0(=3)$ in (d) though the Weyl points penetrating to the bulk becomes absent.
\label{1DDirac}}
\end{figure}
When we approach the pair annihilation of Weyl points, the Fermi arcs on the surface shrink, while the Fermi arcs on the domain walls become elongated as we see in Fig.\ref{1DDirac}. 
After the pair annihilation of the Weyl points,
the surface is no longer a topological boundary.
However, the arc on the domain wall forms a closed loop (open Fermi line connected through Brillouin zone boundaries).
For a given $\kappa$, 
the topological invariant changes its sign at the domain wall, as
the 1D weak Chern insulators. The existence of the ingap state at the domain wall is 
protected by the spatial symmetry preserved around the domain wall.

Though the symmetry is not the same, the protection of the ingap state as the topological character of  1D effective models has a similarity to the polyacetylene mentioned above. The present domain wall may be regarded as a 3D extension of the polyacetylene soliton.

\subsection{ Solution of three-dimensional model }
\label{Section6}
\begin{figure}[htb]
\centering
\includegraphics[width=4.5cm]{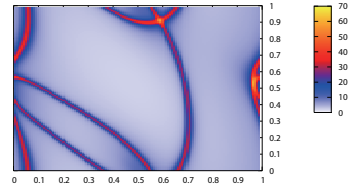}
\caption{
(Color online) Spectral function projected to domain wall in $(01\overline1)$ direction for $U/t=4$, $\zeta/t=-0.2$,
and $k_{\rm B}T/t=0.1$ with a Lorentzian width $\delta=0.01t$.
At this temperature,
bulk Weyl points do not exist any more.
Sharp curves of high spectral weight indicates the Fermi surface.
\label{3DFermiSurface}}
\end{figure}

The essential properties of the simple 1D
Dirac equation in the previous section
is preserved in the fully unrestricted Hartree-Fock
analysis of eq.~(\ref{Hamiltonian}).
The solution of the large supercell calculations for a typical $(01\overline{1})$
domain wall with the parameters $U/t=4$ and $\lambda/t=-0.2$
is illustrated in Fig.~\ref{3DFermiSurface}
The arcs on the surface and the domain wall appear in the complementary side of the Weyl points.  With the growth of the ordered magnetic moment, the Weyl points are annihilated in pair and the Fermi arc transforms to the full Fermi line at the domain wall as we see in Fig.~\ref{3DFermiSurface}. 

A remarkable property of the domain wall is not only the gapless metallic excitations but also that 
it bears 
uniform magnetization
perpendicular to the domain-wall plane.
More remarkably, the amplitude of the magnetization per unit area of the domain wall
is a constant irrespective of the domain wall direction and only determined by $m$
within numerical errors.
Therefore,
the total magnetization of a closed domain wall surrounding a domain 
will vanish because of the cancellation between the opposite side of the domain, if the underlyng all-in-all-out order has a uniform order parameter $m$.
The cancellation of the magnetizations is analogous to that of a pair of spin soliton and anti-soliton of polyacetylene.\cite{SuSchriefferHeeger}

However, the cancellation is not perfect, if $m$ has spatial inhomogeneity arising from  external origins such as electric fields, defects, lattice strain, charged impurities, and doped carriers. The net magnetization may appear even by the self-doping or spontaneous distortion around the domain wall, which is stabilized under the magnetic field so as to gain the Zeeman energy. 
This
is essentially inverse effects of magneto-strain
and magneto-charge responses.\cite{Arima13}

\subsection{Comparison with experimental results of pyrochlore iridates}
\label{Section7}
Several puzzling experimental results have been reported for the pyrochlore iridates R$_2$Ir$_2$O$_7$ when they show the all-in-all-out type magnetic order:
\begin{enumerate}
\item[1)] The DC resistivity does not follow the expectation from the simple antiferromagnetic insulator, because the resistivity rather follows the temperature dependence of the variable range hopping and at least does not follow the activation type, indicating ``bad insulators".\cite{Matsuhira11,Ishikawa,Ueda12}  
\item[2)]  The field cooled and zero-field cooled samples show different magnetic responses and the field cooled sample shows weak ferromagnetic moment in the order of $10^{-3} \mu_{\rm B}$ per unit cell.\cite{Matsuhira11,Ishikawa}  This is again not expected in a simple antiferromagnetic insulator.
It is even more puzzling because the polycrystal shows weaker magnetization than the single crystal.\cite{Ishikawa}
\item[3)] When R is a magnetic ion like Gd or Nd, they show very large negative magnetoresistance.\cite{Matsuhira13}   
\end{enumerate}
Let us discuss what is expected from the present domain wall theory.
By cooling,
magnetic domain walls are inevitably introduced.
The gapless states around the domain wall contribute to the conduction even when the Weyl points are all annihilated. This explains the robust conduction at low temperatures with the well developed all-in-all-out order.  In addition, the doubly degenerate gapless states, one coming from the $m>0$ side and the other from $m<0$ are spatially formed in the opposite side of the domain wall, but the distance is the order of the unit cell particularly if $|m|$ gets large.  Therefore, the overlap of the two degenerate wavefunctions is nonzero and electrons can be scattered between these two states. This is a point distinct from the surface state of the topological insulator, where the surface state does not have back scatterings and the anti-localization effect exists. Because of the two-dimensionality of the domain wall, this overlap causes weak but nonzero Anderson localization.  This explains the robust but bad conduction in the experiments.  To keep away from the back scatterings, it is important to lift the degeneracy,
which is left for future studies.

The difference in the zero-field and field cool samples is explained by the fact that the domain walls are  pinned at their favorable impurity/disorder sites under the magnetic field 
to optimize the net magnetization along the
external magnetic fields. 
The self-doping or self-distortion may also occur to lower the Zeeman energy coming from the magnetization around the domain wall.
This induces a nonzero magnetization only under the magnetic fields.
A realistic value of self doping $n_{\rm ex}\sim 10^{-3}$ explains the peculiar uniform magnetization ($\sim 10^{-3}\mu_B$/unit cell) universally observed experimentally.\cite{Matsuhira11}
The smaller magnetization for polycrystals~\cite{Ishikawa} is consistent because magnetic domains are wiped out. The larger hysteresis for stoichiometric samples~\cite{Ishikawa} is simply ascribed to stronger all-in/all-out order. 
Strong sample dependence~\cite{Ishikawa} and hysteresis in the magnetization sweep~\cite{Matsuhira13} at lowest temperatures also support this view.

A tempting explanation of the large negative magnetoresistance for Nd/Gd compounds~\cite{Matsuhira13,Disseler} is the 
fluctuating ferromagnetic moment of Nd or Gd induced by the magnetization at the domain wall at zero field,
which scatters carriers at domain walls similarly to the double-exchange mechanism.
Under the magnetic fields, this fluctuation is suppressed and reduces the scatterings,
which contributes to the large negative magnetoresistance.

\section{Concluding remark and future perspective}
The topological Mott insulator stabilized only by the electronic Coulomb correlation rather than the spin orbit interaction offers a unique way of controlling the topological nature of electronic systems because the spontaneous symmetry breaking of the topological state can be switched on and off by external parameters such as temperature, pressure and magnetic field. It is also useful because we do not need to use the heavy rare elements with strong spin-orbit interactions.
 
The quantum criticality of the transition to the topological Mott insulator does not follow the conventional Landau-Ginzburg-Wilson scheme, because the symmetry breaking is combined with the topological change of the Fermi surface, which is not anticipated in the Landau-Ginzburg-Wilson scenario.  Because the critical exponents suggest ``soft" transitions, we expect larger fluctuations that may also induce novel quantum phases.

It has turned out that the magnetic domain wall of the pyrochlore iridates offers an interesting possibility of the control of the electronic conduction by the magnetic fields through the domain wall conduction. This mechanism may provide a potential application for the magnetic control of the transport. It is desired further to understand them more quantitatively for a better magnetic control. 

\begin{acknowledgements}
The series of the work was financially supported by MEXT HPCI Strategic Programs 
for Innovative Research (SPIRE) (with the grant number hp130007) and Computational Materials Science Initiative (CMSI).
This work was also supported by Grant-in-Aid for 
Scientific Research {(No. 22104010, No. 22340090, and No. 23740261)} from MEXT, Japan.
\end{acknowledgements}

\end{document}